\documentclass{elsart}
\usepackage{natbib}
\usepackage[dvips]{graphicx}
\begin{document}
\begin{frontmatter}

\title{The Cosmic Ray Muon Flux at WIPP}

\author[lanl,jlu]{E.-I. Esch},
\ead{ernst@lanl.gov}
\author[lanl]{T.J. Bowles}, 
\author[lanl]{A. Hime}, 
\author[lanl]{A. Pichlmaier\thanksref{ca}},
\author[lanl]{R. Reifarth}, 
\author[jlu]{ H. Wollnik}
\address[lanl]{Los Alamos National Laboratory,Los Alamos, NM 87545}
\address[jlu]{II. Physikalisches Institut, Justus Liebig Univerit\"at Gie\ss{}en, Gie\ss{}en, Germany}
\thanks[ca]{Current address: Paul Scherrer Institut, 5232 Villigen PSI Switzerland }
\maketitle
\begin{abstract}
In this work a measurement of the muon intensity at the Waste Isolation 
Pilot Plant (WIPP)  near Carlsbad, NM, USA is presented. WIPP is a salt mine with a depth of 655~m.
The vertical muon flux was measured with a two panels scintillator coincidence setup
 to $\Phi_{vert}~=~(3.10^{+0.05}_{-0.07})~10^{-7}$~s$^{-1}$cm$^{-2}$sr$^{-1}$.
\end{abstract}
\begin{keyword}
muon flux \sep muon \sep scintillator \sep underground \sep atmospheric \sep cosmic
\PACS 
96.40.Tv      
\sep 29.40.Mc 
\sep 95.55.Vj 
\sep 95.85.Ry 

\end{keyword}
\end{frontmatter}

\section{Introduction}
The cosmic ray background consists of several different particles such as
pions, electrons, protons, etc. They are coming either from space
or are generated in the high atmosphere. 
 The most numerous generated underground
are muons that are created when protons and secondary particles
interact with atoms in the atmosphere. Experiments can be shielded
from this cosmic activity by going deep underground. In this paper we
describe a measurement of the cosmic ray muon flux at the Waste
Isolation Pilot Plant (WIPP) with an overburden of 655 m of rock and
salt and located 30 miles east of Carlsbad, New Mexico \cite{DOE01}. Such
measurements are useful in assessing the appropriateness of WIPP as a
site for underground experiments that require shielding from cosmic
radiation. The depth requirements and adequacy of a particular site
vary depending on the nature and goals of a particular experiment. This paper
does not attempt to address the appropriateness of the WIPP site as a
general purpose underground laboratory, an exercise of intense
activity within the scientific community at present. Such a task is
beyond the scope of this paper. The paper focuses on the measurements of
the muon flux at WIPP, which serves as an important input in evaluating
the impact on specific experiments.


\section{Experimental Setup} 
The experiment to measure the muon flux in the underground of WIPP
was conducted 655 meters below the surface in the experimental Q-Area. 
The detector consisted of two plastic scintillator
panels each 305~cm long, 76.2~cm wide and 2.54~cm thick. Each panel
had two light-guides mounted on the two narrow ends.  A 12.7~cm diameter
photo-tube with an operating voltage of 800 volts was mounted on each
light-guide enabling a double ended readout. 
The muon detection efficiency as a function of position  
for each panel was measured by placing two small scintillators above and below.
If the small detectors triggered coincident, a muon must have passed through the
large panel. The number of events in the panel over the 
total number of triggers gives therefore the muon  detection efficiency.
Each position was tested with 10,000
coincidence events. 
High energy, minimum ionizing muons would deposit about 5~MeV when
traversing a scintillator paddle at normal incidence. The actual
energy spectrum deposited in the Patel exhibits a broad tail around
this pronounced peak due to muons traversing at angles other than
normal, straggling effects, and the finite energy resolution of the
scintillator. Nonetheless, by summing the signal heights from the
PMT's of each panel with a threshold of 3 MeV, a clean muon signal
can be obtained well above the electronic noise and low energy gamma
background associated with a single PMT channel.
The detector efficiency for muons was found to be~100$\pm1$\% for all positions.

\begin{figure}[h]
  \begin{center} 
  \includegraphics[width=\linewidth]{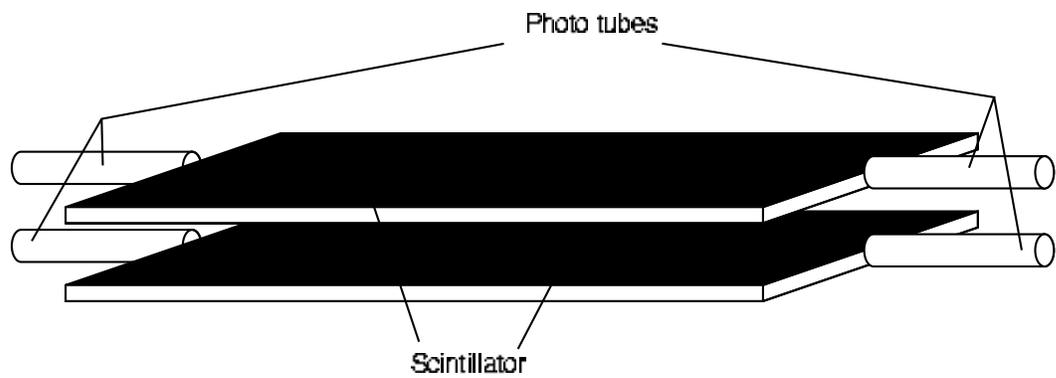}
  \caption[Setup of the scintillator panels.]
  {Setup of the scintillator panels. They were placed on top of
  each other with a distance of 30.5 cm. \label{setup}}
  \end{center}
\end{figure}

\begin{figure}[h]
  \begin{center} 
  \includegraphics[width=\linewidth]{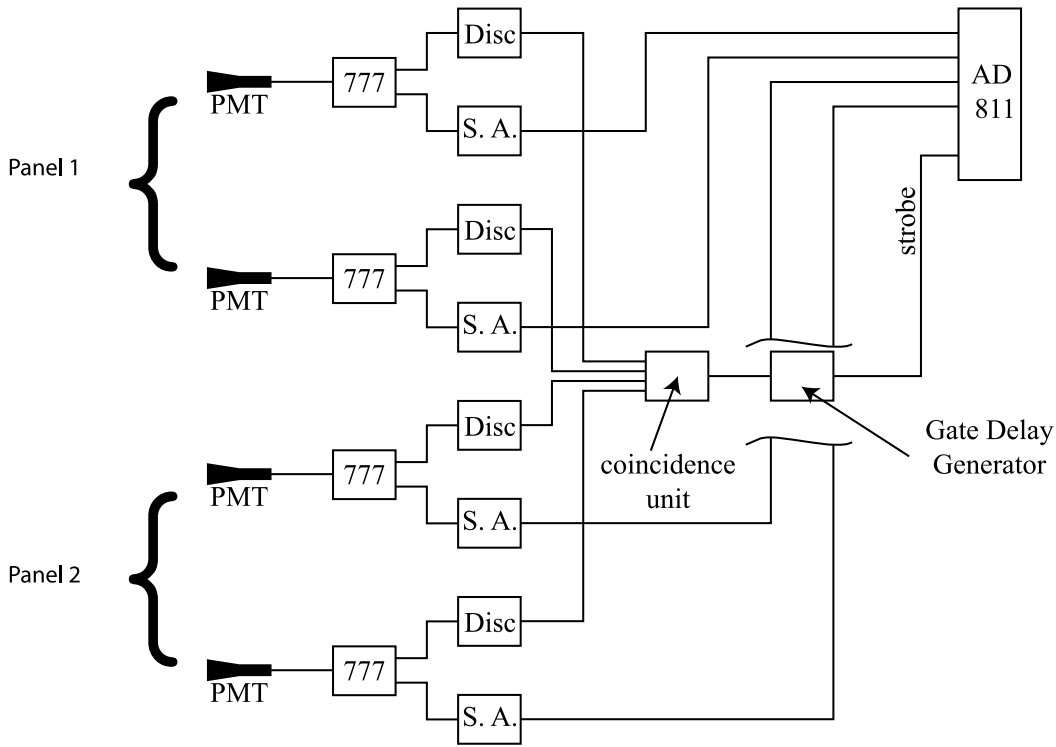}
  \caption[Setup of the scintillator-electronics.] 
  {Setup of the scintillator-electronics. The electronics was setup 
  in a quad coincidence mode. The signal coming from the photo-multiplier-tube
  (PMT) was amplified by a Phillips Scientific fast amplifier (777) with a gain 
  of 20. This output signal was split. One signal was used as input for a 
  discriminator
  attached to a coincidence unit to form the strobe for the LeCroy AD811
  ADC. The other output was run through a shaping
  amplifier (S. A.) into the input channels of the ADC.  
  \label{electronicsetup}}
  \end{center}
\end{figure}

\begin{figure}[h]
  \begin{center} 
  \includegraphics[width=\linewidth]{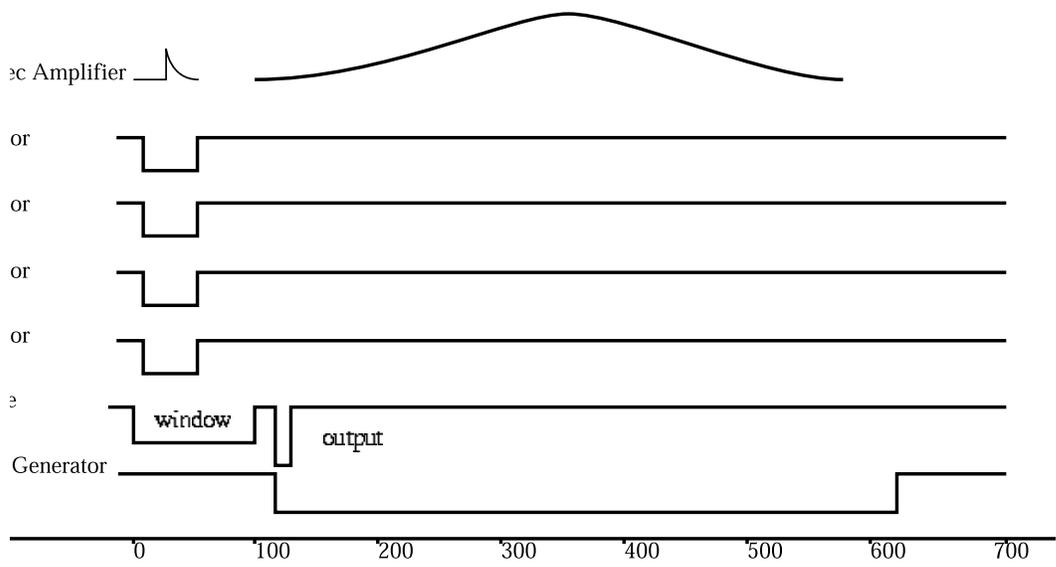}
  \caption{Timing of the scintillator Coincidence.} 
  \label{mtime}
  \end{center}
\end{figure}

For the actual experiment 
panel 1 was mounted on top of  panel 2 in a distance of
30.5~cm (see Fig.~\ref{setup}). A fourfold coincidence from the photo-tubes 
was required
to record an event.  The high voltage on each
photo-multiplier tube (PMT) was adjusted by demanding an equal pulse height   
 for a $^{90}$Sr calibration source place in the center of each panel.  
The decay product of $^{90}$Sr is  $^{90}$Y which decays emitting 
high energy electrons, which were detected in the panel.   
Fig.~\ref{electronicsetup} shows the electronic diagram.
The pre-amplifier signal had a rise
time of 10~ns to 20~ns. It was split up and one
signal was fed through a spectroscopic amplifier with a shaping time
of 0.25~$\mu$s and connected to an LeCroy AD811 ADC.  The other signal was put
into a discriminator with a 175~mV threshold which was equivalent to an
energy of 1.1~MeV.  The steep NIM pulses from the discriminators
were used as input for a coincidence unit. Fig.~\ref{mtime} illustrates the 
timing for the different components during an event.
The
coincidence window had a width of 100~ns. The output of
the coincidence was then processed by a gate/delay generator and used
as strobe for the ADC. The strobe window was chosen to be 0.5~$\mu$s
wide. This ensured that the peak from the spectroscopic amplifier was
inside the ADC strobe window.
\section{Data Analysis} 
The experiment was conducted during a run-time of 6.17 days. 
Panel one had a slightly better resolution than panel two.
This was due to an increased noise in photo tube 3.
The total charge in a panel was calculated as
the sum of the outputs of its two PMTs. 

\begin{figure}[h]
  \begin{center} 
  \includegraphics[width=\linewidth,clip=true]{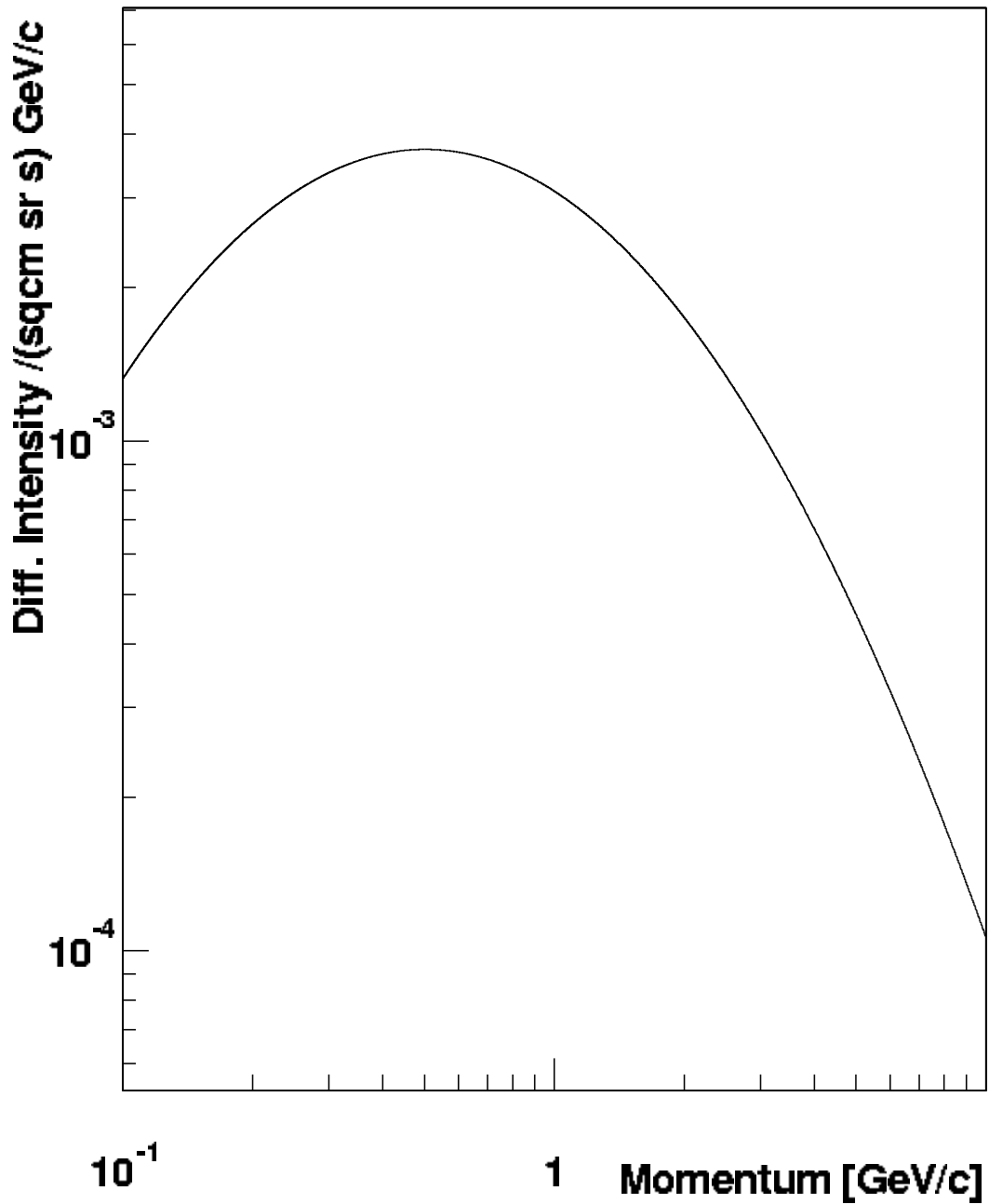} 
  \end{center} 
  \caption {\label{muonenergy}Differential muon intensity versus 
  muon momentum.The spectrum is an
  empirical fit of the momentum distribution of vertical muons at sea level.
  The spectrum has a 
  flat maximum at about 0.5 GeV/c. }
\end{figure}

\begin{figure}[h]
  \begin{center} 
  \includegraphics[width=0.8\linewidth]{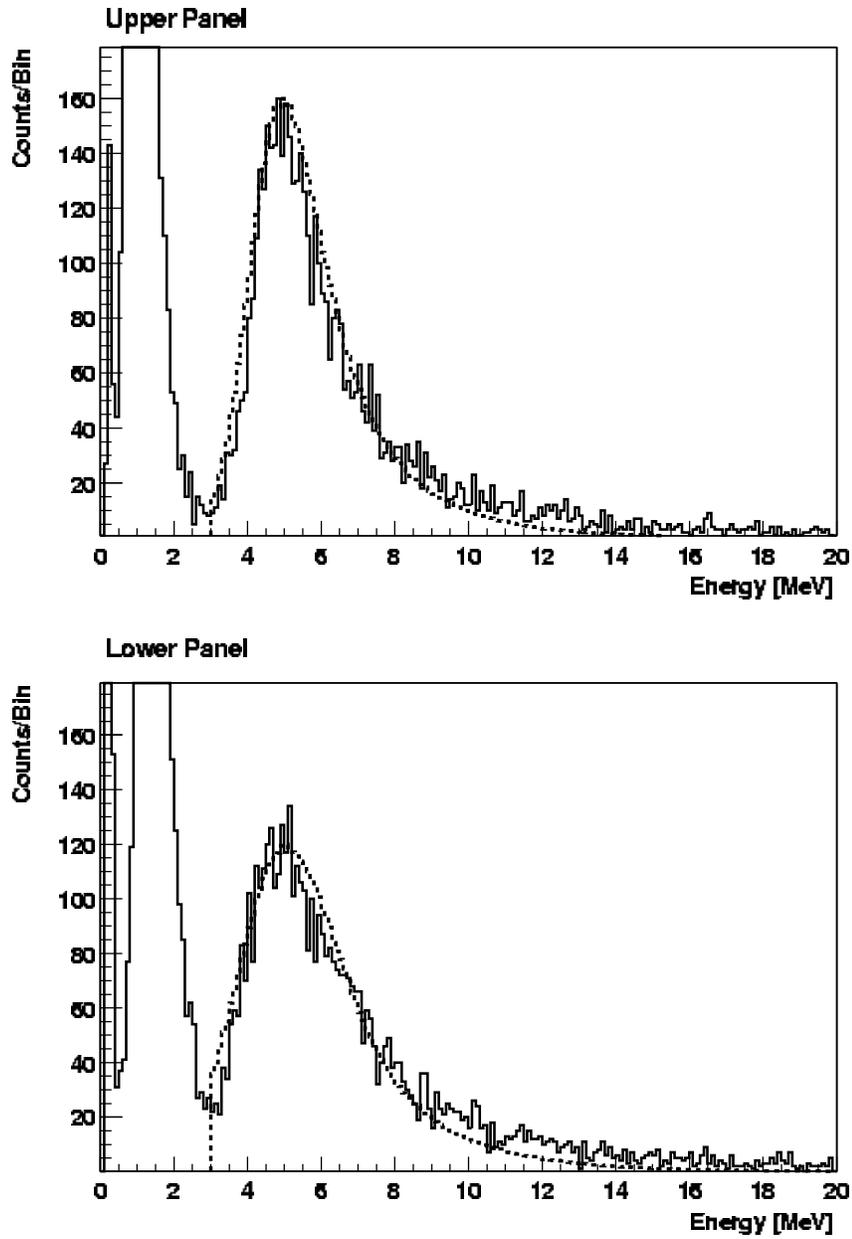}
  \caption[Spectrum for the raw energy in the two scintillator panels.] 
  {Spectrum for the raw energy in the two scintillator panels. 
  The energy is plotted into a histogram  in units of ~MeV, the 
  Monte-Carlo simulation is shown as dotted line. \label{rawenergy}}
  \end{center}
\end{figure}

To determine the energy calibration of the
detectors Monte-Carlo simulations were carried out with Geant 4. 
The initial energy distribution of the muons simulated was derived from the
following empirical formula corrected for the energy losses in 655~m of rock and salt
\cite{MIY73}:
\begin{eqnarray}  
N(P) dP &=& AP^{-\alpha}\,dP \mbox{ with} \label{wol}\\ 
\alpha &=& 0.5483+0.3977\ln P \mbox{ and} \nonumber \\
A&=& 3.09\times 10^{-3}\mbox{cm}^{-2}\mbox{s}^{-1}\mbox{sr}^{-1}
(\mbox{GeV}/c)^{-1}, \nonumber
\end{eqnarray}
$P$ being in GeV/c. The intensity distribution of the polar angle $\theta$
was taken from \cite{MIY73}.  A simplified angular distribution for medium
depth  can be written as \cite{MIY73}
\begin{eqnarray} 
\Phi(h,\theta)&=&\Phi(h,0) \cos ^{1.53} \theta 
                 e^{-8.0 \times 10^{-4} h(\sec \theta -1)}\label{myaki2}.
\end{eqnarray}
with $\Phi(h,\theta)$ as the 
flux for a specific angle in cm$^{-2}$s$^{-1}$sr$^{-1}$. Figure~\ref{muonenergy} 
shows this spectrum at sea level. 
The energy calibration
was done by fitting the maximum intensity of the muon peak to the maximum
intensity of the Monte-Carlo for the detector.
A comparison  of measured and simulated data is shown in 
Fig.~\ref{rawenergy} for both panels. 

The total muon spectrum for each panel
was achieved by adding the ADC values of both photo tubes up on an
event-by-event base. 
The different energy resolutions of the PMT's were taken into account during the 
Geant~4 simulations (see difference between upper and lower panel in 
Fig.~\ref{rawenergy}). For more details see \cite{ESC01}. 
The shape of the measured spectrum could be reproduced 
within statistical uncertainties. 
The muon peak with its maximum intensity around 4.8~MeV is clearly visible. 

\begin{figure}[h]
  \begin{center} 
  \includegraphics[width=\linewidth]{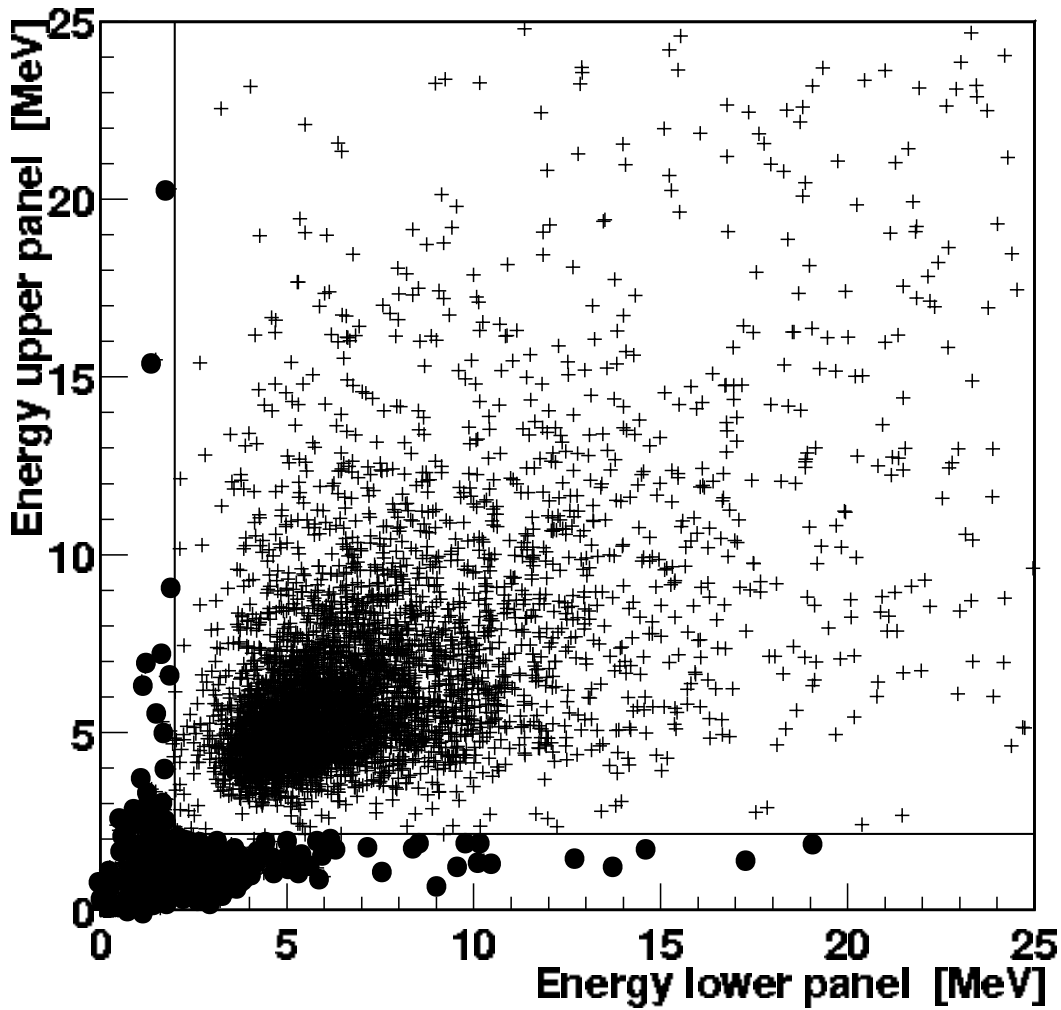}
  \caption[Two dimensional spectrum for the cuts applied to the muons.] 
  {Two dimensional spectrum for the cuts applied to the muons. The
  spectrum shows on the x-axis the deposited energy in MeV for the lower
  panel, on the y-axis the deposited energy for the upper panel. The
  black line displays the cut chosen for the muon number. The events marked by
  a cross
  represent the muon peak, the events marked by a dot are dominated by the 
  random
  coincidence of the PMTs.
  \label{muoncut}} 
  \end{center}
\end{figure}

\begin{figure}[h]
  \begin{center} 
  \includegraphics[width=\linewidth]{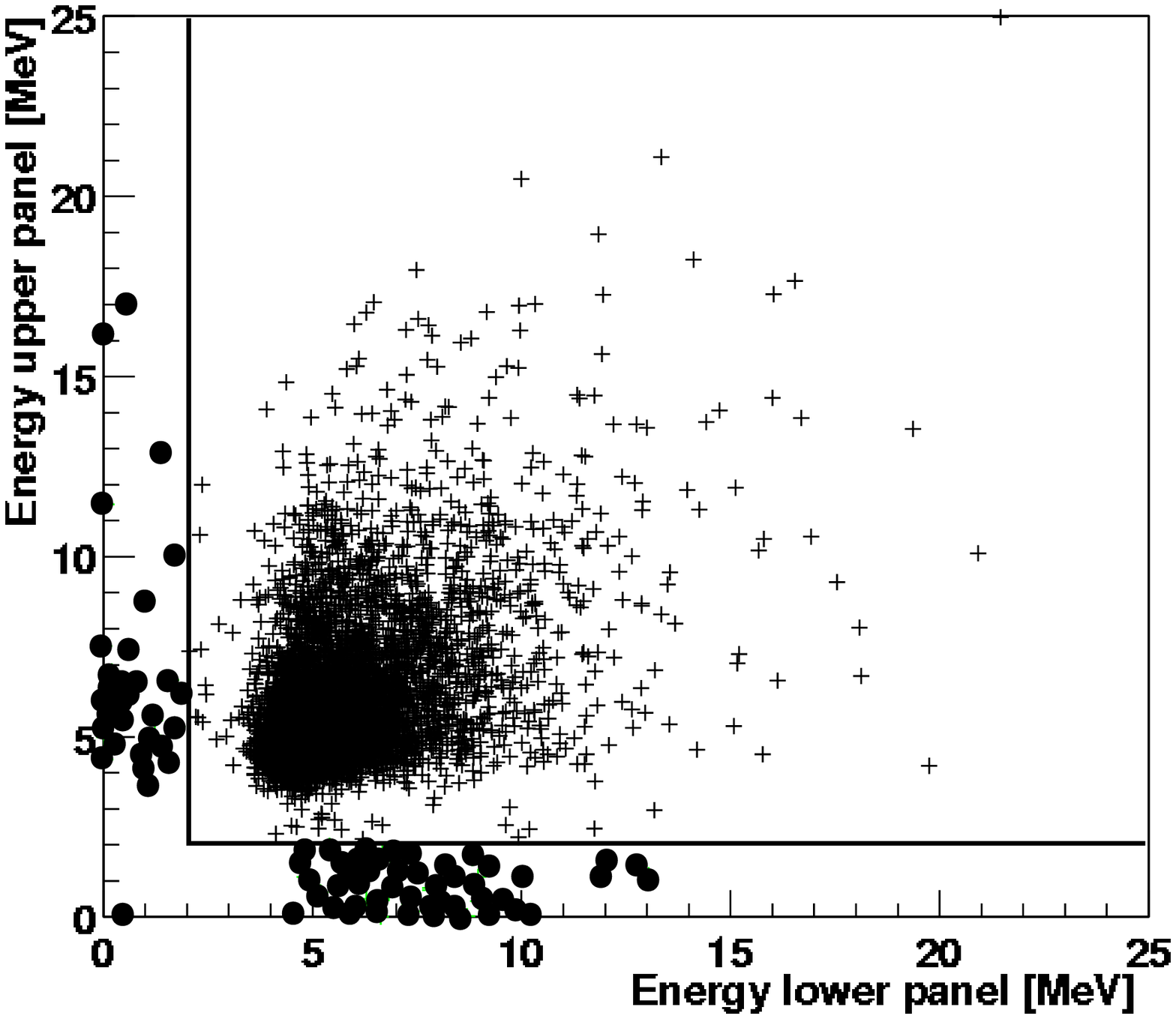}
  \caption[Two dimensional spectrum for the cuts applied to the muon
  Monte-Carlo simulation.]  {Two dimensional spectrum for the cuts applied to the
  muon Monte-Carlo simulation. The spectrum shows on the x-axis the deposited energy
  in MeV for the lower panel, on the y-axis the deposited energy for
  the upper panel. The black line displays the cut chosen for the muon
  number. The events marked by a cross represent the muon peak, the 
  events marked by a dot represent 
  muons outside the cut region. The ratio of these muons in the simulation
  is $1.24\pm0.01\%$.\\
  \label{mccut}} 
  \end{center}
\end{figure}

Fig.~\ref{muoncut} shows a two dimensional plot of an event by
event energy distribution for the two panels. The x-axis shows the energy
deposition in the lower panel the y-axis the energy deposition in the upper panel.
Among the events marked with a cross,
the muon peak is clearly visible at approximately 5 MeV energy deposition in
each panel.  
A cut above 2~MeV energy
deposition for each panel was chosen to determine the number of muons.
The black line represents the cutting border. 
Fig.~\ref{mccut} shows the same cut applied to the Monte-Carlo simulations.
According to the Monte-Carlo simulations  $1.24\pm 0.01\%$ of the muon events
are cut out. 
The low energy events in Fig~\ref{muoncut} are due to PMT noise, which are
therefore absent in Fig.~\ref{mccut}. 
Table~\ref{cuts} contains the integral information of the different regions in
the spectrum shown in Fig~\ref{muoncut}.
The errors quoted are
purely statistical errors.  The final muon peak cut is shown by cut 6
and the muon number derived from the cut is 
\begin{eqnarray}
N_\mu &=& 5202\pm72,
\end{eqnarray}
where the uncertainty is statistical only. 

The next step is to determine how many counts of the background
contributed into the signal and how many counts from the signal were lost
into the low energy background due to the cut. 
The trigger rate coming from two photo tubes in one panel in coincidence was
220$\pm$10~Hz. As described above, during the experiment the data were taken demanding 
a quadruple coincidence, which basically means a coincidence between the two
panels. The random coincidence rate can therefore be calculated with 
\begin{eqnarray}
\nu_r&=& 2\nu_1\nu_2 \tau,
\end{eqnarray}
where $\nu_r$ is the random coincidence frequency, $\nu_1$ and $\nu_2$ the
two frequencies and $\tau$ the coincidence window. With
a coincidence window of $100\pm 5$~ns the rate can be calculated to
$\nu_r=9.68 \pm0.79\times 10^{-3}\mbox{ Hz}$ or 5158 $\pm 43$ counts.
The main part of the background can be attributed to random
coincidences (see first cut 0 of Table~\ref{cuts}) 
the tail of the background was assumed to be exponential and extrapolated 
into the region of interest. The results are listed in Table~\ref{tails}.
The number of muon induced events outside the region of interest 
was estimated using the result of the Monte-Carlo simulations (see 
Fig.~\ref{mccut} and Table~\ref{cuts}). 

The muon flux can
be calculated as follows.
\begin{eqnarray}
\Phi_{total} &=& \frac{N_0}{\epsilon \cdot a \cdot t_{rt}} \label{ntot},
\end{eqnarray}
where $\Phi_{total}$ is the total flux of muons through the panel,
$N_0$ the measured number of muons, $\epsilon$ the efficiency of the detector
which includes both the physical efficiency of the scintillator and the
geometrical efficiency due to the setup of the detector and the angular
distribution of the muons, $a$ the area covered by the detector (2.33~m$^2$) 
and $t_{rt}$ the run-time of the experiment which was
532800 seconds.

The deviation of the geometrical efficiency from 100\% results from the fact that 
not all muons pass the upper plate vertically. By requiring a coincidence between 
the upper and the lower panel, the measured number of muons will therefore be less than
the number of muons passing through the upper panel. The deviation of the geometrical 
efficiency was determined by simulating 
two surfaces with size of the scintillator panels with a separation of 30.5~cm.
The start positions for the muons were randomly chosen on the upper panel. 
The azimuthal direction was sampled randomly in an
interval between 0 and 2$\pi$. The simulations were carried out with a 
lateral distribution according to equation~\ref{myaki2}. The depth parameter h was
varied over a range of $\pm$10\%. 
With a measured density of (2.3$\pm$0.2)~g/cm$^3$ 
\cite{LAB95} an efficiency of (88.5$\pm0.2$)\% was obtained.
The vertical flux can be calculated using the approximating formula from
\cite{MIY73}
\begin{eqnarray} 
\Phi(h,\theta)&=&\Phi(h,0) \cos ^{1.53} \theta \label{thetaref},
\end{eqnarray}
for a depth of 1526 m.w.e. the vertical flux is 
$\Phi_{vert}=\left(0.65\pm0.04\right)\times \Phi_{tot}$.

\subsection{Backgrounds}
In the following section other sources of background than discussed in previous 
sections are estimated. In order
to generate a coincidence pulse with an energy deposition of more than
the threshold of 2.0~MeV one has to look at particles with higher energy. 

\subsubsection{Neutrons from Muons}
Neutrons generated from muons traversing through the rock and salt are one 
candidate for the expected background in the WIPP.
In his paper Bezrukov \cite{BEZ73} estimates the
amount of neutrons per muon generated at a depth of
1500 meter water equivalent (m.w.e.) in hg/cm$^2$ to
\begin{eqnarray}
N_{n(\mu)} &=& 7~10^{-4} \frac{\mbox{cm}^2}{\mbox{g}}.
\end{eqnarray}

\begin{figure}[h]
  \begin{center} 
  \includegraphics[width=10cm]{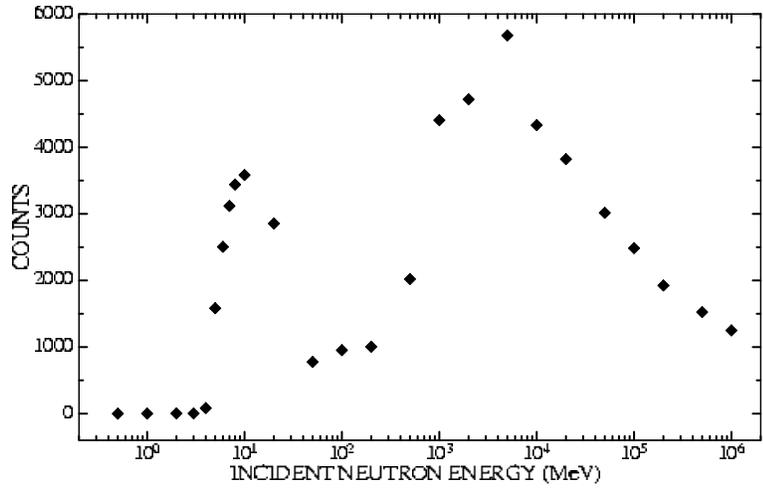}
  \caption[Neutron efficiency.]  {Neutron efficiency of the two panels as a 
  function of primary neutron energy.}
  {\label{neff}}
  \end{center}
\end{figure}

With equation the empirical equation from \cite{MIY73}
\begin{eqnarray} 
\Phi(h,0)&=&\frac{174}{h+400}\left( h + 10\right)^{-1.53}
            \mbox{e}^{-8.0 \times 10^{-4} h},\label{myaki1}
\end{eqnarray} 
where $\Phi$ has the units of (cm s sr )$^{-1}$ and h represents the depth in
$hg/cm^2$,
this calculates to a neutron flux of 
$3.36$ $10^{-7}~\mbox{kg}^{-1}\mbox{s}^{-1}$.
With an assumed average cross section of 1~barn the attenuation length in salt
is 21~cm. This calculates to a neutron flux of  
$1.6$~$10^{-8}~{\mbox{cm}^{-2}\mbox{s}^{-1}}$, 
or $8.6$~$10^{-3}~\mbox{cm}^{-2}$ during the entire length of experiment of 6.17 days. 
Several Monte-Carlo simulations have been carried out to estimate the detection efficiency for neutrons
in the existing detector geometry. The neutrons were randomly started  from a sphere of 2~m
radius surrounding the two panels into the whole solid angle. Several simulations with neutron
energies ranging from 500~keV to 1~TeV were performed (see Fig.~\ref{neff}). According to the experimental 
settings, only such events were counted,
where both panels had an energy deposition between 2 and 20~MeV. The most important
interaction mechanism is inelastic scattering on hydrogen and carbon in the scintillator material. Therefore
the number of events depositing at least 2~MeV in each of the panels is reduced by at least 3 orders of
magnitude, if the incident neutron energy falls below 4~MeV. Only neutron captures and other nuclear 
channels can than produce the required energy.  
The number of simulated events in Fig.~\ref{neff} corresponds to 20~neutrons/cm$^{2}$. 
Scaling this number to the expected number of neutrons during the experiment and assuming 
the highest efficiency, gives an upper limit for the number of neutron induced events of $N_{neutron}~=~2.6$.
Most of the neutrons will be moderated by the time they reach the detector. The assumption 
made above is, therefore, very conservative.

\subsubsection{Neutrons from U and Th}
Natural radioactivity is
present in every geological layer of the earth.  An abundance of uranium
and thorium in the salt is able to generate neutrons via ($\alpha ,
n$)-reactions and through spontaneous fission. The U and Th contents at WIPP 
have been measured \cite{WEB98} and are displayed in
Table~\ref{uandth}. With the numbers from \cite{FLO88} for neutron production via ($\alpha,n)$-reactions 
of U and Th in salt, who lists  U$_{(\alpha,n)}$=5.07~10$^{-8}$~s$^{-1}$g$^{-1}$ 
for an uranium abundance of 0.3~ppm and Th$_{(\alpha,n)}$=1.51~10$^{-7}$~s$^{-1}$g$^{-1}$ for a 
thorium abundance of 2.06~ppm in salt, 
one can estimate the neutron production rate to
U$_{(\alpha,n)}$=8.11~10$^{-9}$~s$^{-1}$g$^{-1}$ for an uranium abundance of 0.048~ppm 
and Th$_{(\alpha,n)}$=1.83~10$^{-8}$~s$^{-1}$g$^{-1}$ for a thorium abundance of 0.25~ppm
in salt (according to column "Used Values" in Table~\ref{uandth}). 
Assuming again an attenuation length of 21~cm leads to an upper neutron flux of 
1.28~10$^{-6}$~s$^{-1}$g$^{-1}$.
Since the highest alpha generated in the U/Th decay chain has ca. 8~MeV and the Q-values of
the (n,$\alpha$) reactions at Na or Cl is at most -3~MeV, a maximum neutron energy 
of 5~MeV can be assumed. 
With the simulation results from Fig.~\ref{neff} this leads to a neutron number in the
panels of $N_{U,Th} = 54$ during the measurement. 
 
\subsubsection{Gamma Background}
There are two main sources of gamma background, first the decay of U/Th, and second the
decay of $^{40}$K.
The highest level of gamma background comes from the decay of $^{40}$K, which is abundant
in the surrounding salt.
The abundance of potassium in salt is 784~$\mu$g/g 
(see Table~\ref{uandth}) and corresponds to $1.5~10^{15}$~$^{40}$K/g.  
A 1.46~MeV $\gamma$-ray will be emitted during the decay, with a probability 
of 10.7\%. 
With an half-life of
$t_{1/2}=1.28~10^9$~y this results in a decay rate of
approximately 2.7~10$^{-3}$~s$^{-1}$g$^{-1}$. 
With the macroscopic scattering cross section from \cite{BER01} of $\sigma$=$5.02~10^{-2}$~b 
in NaCl and the assumption of an exponential attenuation law,  
an attenuation length of 8.4~cm is computed. The attenuation length is defined as
the length after which the intensity drops by a factor e.
With a salt density of 2.3~g/cm$^3$
the amount of salt contributing $\gamma$-rays can then
be estimated to 19.3~g/cm$^{2}$, which results
in a $\gamma$-flux of $\Phi_{\gamma}$~=~5.22~10$^{-2}$~s$^{-1}$cm$^{-2}$.
Since each gamma can deposit at most 1.46~MeV, a
triple coincidence during the coincidence
window of 100~ns in the detector is required to generate a signal 
inside the muon-cut region. 
According to the GEANT simulations, a $\gamma$-ray between 1 and 2 MeV emitted from a sphere
of 2~m radius around the panels deposits energy with a probability less than 1\% in one of the detectors,
resulting in a rate of 262~s$^{-1}$ energy-depositing events. The resulting rate of triple
coincidences ($\tau^2\nu^3$ with $\tau$~=~100~ns and $\nu$~=~262~s$^{-1}$) is 1.80~10$^{-7}$~s$^{-1}$,
or 0.1~events during the experiment, hence negligible.

The number of gammas originating from $(n,\alpha)$ and $(n,\gamma)$ reactions can be estimated 
using the above calculated neutron flux. Assuming that only one $\gamma$-ray 
above 5~MeV is produced per neutron, the flux is 1.28~10$^{-6}$~s$^{-1}$g$^{-1}$.
Gamma rays with energies up to 10~MeV have been simulated in order to estimate the
efficiency of the muon panels including the coincidence requirement of at least 2~MeV 
energy deposition per panel. The geometry and 
number of particles were the same as described above for the neutrons (see Table~\ref{nsim}). 
Assuming that the detection efficiency was on average below the simulated efficiency for a
10~MeV $\gamma$-ray, at most of 32 counts are expected during the entire 
run of the experiment.  
 
\subsection{The Muon-Flux}
The raw data have (5202$\pm$72) counts in the region of interest. 
After the correction for electronic noise inside the region of interest (40$\pm$4) events 
and muon-induced events outside this region (62$\pm$6) events one obtains
(5224$\pm72\pm7$) counts. A conservative 10\% uncertainty was assumed for each of the corrections,
which were then quadratically combined (see Table~\ref{factors}).  
Only upper limits for the contribution from neutrons (55) and $\gamma$-rays (32) have been performed.
The uncertainty due to the detection efficiency is ($\pm$55) events.  
Since all the uncertainties are independent from each other, a quadratical addition of the
different components can be performed.
The final number of muon induced events in the muon panels is therefore ($5224^{+89}_{-108}$).
As discussed above, the geometrical efficiency, resulting from the fact that not all muons are 
vertically, has to be taken into account. The number of muons passing through the detector area
is ($5902^{+102}_{-123}$) events. The uncertainties of both, the surface area of the detector 
and the running time is well below the 1\% level and can therefore be neglected.

Now the muon flux can be calculated:  
\begin{eqnarray}
\Phi_{tot}&=& \frac{ 5902^{+102}_{-123}}
{532800\,s \times 23225 \, \mbox{cm}^2} \nonumber \\
 &=& (4.77^{+0.08}_{-0.10}) \times 10^{-7}\, \, \frac{\mbox{Hz}}{\mbox{cm}^2}
\end{eqnarray}
  
\begin{figure}[ht]
  \begin{center} 
  \includegraphics[width=10cm]{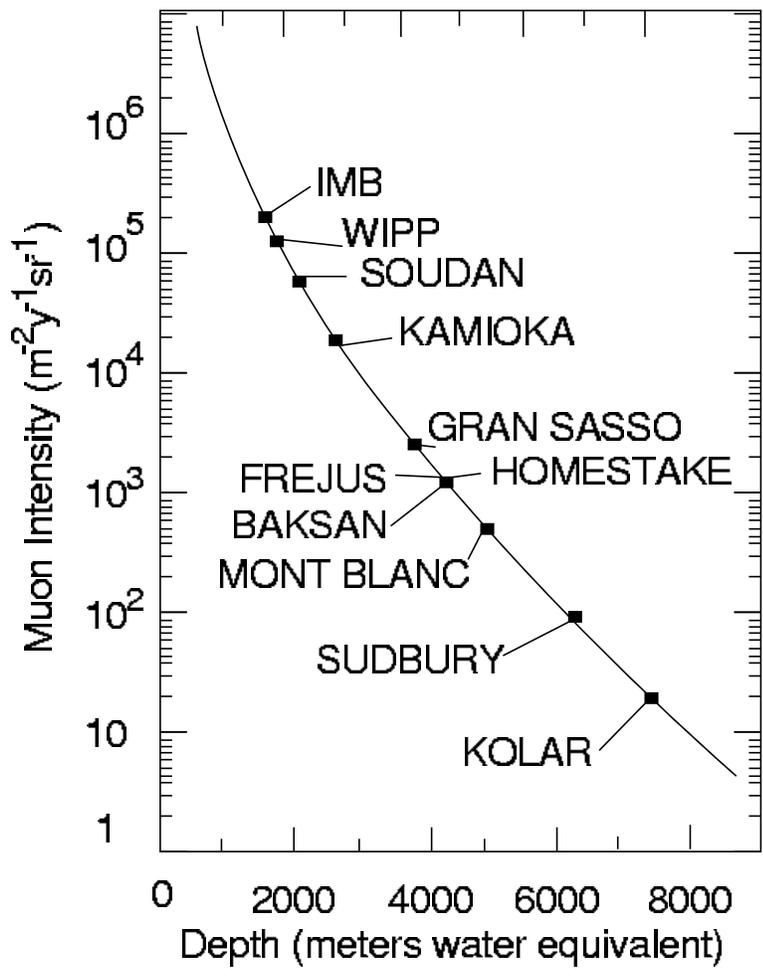}
  \caption[Vertical muon flux for different underground experiments.] 
  {Vertical muon flux for different underground experiments. The errors are
  smaller than the data marker.
  \label{muonflux}} 
  \end{center}
\end{figure}

This flux can now be converted to a vertical flux 
of
\begin{eqnarray}
\Phi_{vert} &=& (3.10^{+0.05}_{-0.07}) \times 10^{-7} \, \, 
\frac{\mbox{Hz}}{\mbox{cm}^2\, \mbox{sr}} ,
\end{eqnarray}
converting to a meter water equivalence of ($1585^{+11}_{-6}$)~m.
Measurements made in the past in different underground laboratories in
the world \cite{CRO87}, \cite{AMB95} and \cite{AND87}
show similar results for this depth. The shallow depth agrees within
its error bar with a fit through the previous experiments.  
Fig.~\ref{muonflux} shows WIPP in comparison to other
underground laboratories. With its average density of 
(2.3$\pm$0.2)~g/cm$^3$
and a depth of 655~m our measurement fits very well
with the predicted data.

\section{Conclusion}

The work described here has determined the vertical muon flux at the WIPP mine to 
$\Phi_{vert}~=~(3.10^{+0.05}_{-0.07})$~$10^{-7}$~s$^{-1}$cm$^{-2}$sr$^{-1}$. This
result agrees well with measurements at other underground laboratories and can now serve
as an experimental basis for low-background experiments.

\section{Acknowledgments}
We would like to thank the WIPP staff, especially Roger Nelson, Dennis Hoffer and Dale Parish
for their excellent support and generous help.
We also want to acknowledge the help of M. Anaya and W.
Teasdale for their technical support. 
This work was partially funded by LDRD funds from LANL and DOE. 

\pagebreak
\begin{table}[t]
\caption{Cuts applied to determine event numbers in muon peak.\label{cuts}}
\begin{center}
\begin{tabular}{c|c|c|r|r}
Cut & Cut & Cut & Events & Count rate \\
Number & Upper & Lower &  & [$Hz$] \\
 &Panel & Panel &   &  \\
\hline
0& E$<$2~MeV & E$<$2~MeV & 5207$\pm$ 72 &  9.8$\pm$0.14 $ \times 10^{-3} $\\
1& --- & --- & 11226$\pm$ 106 &  2.11$\pm$0.02 $ \times 10^{-2} $\\
2& E$<$2~MeV & --- & 5881$\pm$77 &  1.10$\pm$0.014 $ \times 10^{-2} $  \\
3& --- & E$<$2~MeV & 5350$\pm$73 & 1.00$\pm$0.014 $ \times 10^{-2} $ \\
4& E$>$2~MeV & --- & 5345$\pm$73 & 9.7$\pm$0.14 $ \times 10^{-3} $ \\
5& --- & E$>$2~MeV & 5876$\pm$77 & 9.8$\pm$0.14 $ \times 10^{-3} $\\
6&  E$>$2~MeV & E$>$2~MeV & 5202$\pm$72 & 9.9$\pm$0.14 $ \times 10^{-3} $ \\
\end{tabular}
\end{center}
\end{table}
\begin{table}[b]
\caption{Tail fits of the background and muon signal. \label{tails}}
\begin{center}
\begin{tabular}{l|r@{--}l|c}
Tail Name & \multicolumn{2}{c|}{Fit Interval} & counts \\ 
\hline
Monte-Carlo cut off & 0&2.0~MeV & 62 \\
background upper panel & 1.2&2.1~MeV & 20 \\ 
background lower panel & 1.0&2.4~MeV & 40 \\ 
\end{tabular}
\end{center}
\end{table}
\begin{table}[p]
\caption{Natural radioactivity at the WIPP-underground. The
abundance of the elements was measured with mass spectrometry and gamma
spectrometry by \cite{WEB98}. The Averages for the measurements are shown in the first two 
columns. The column labeled Used Value displays the value recommended by \cite{WEB98}. The 
values for the columns in soil were typical measurement of regular ground dirt
gathered in southern New Mexico. The las column displays the ratio of the typical
soil value over the corresponding WIPP number.
} 
  {\label{uandth}}%
\begin{center}
\begin{tabular}{l||c|c|c||c|c|c||c}
& \multicolumn{3}{c||}{at WIPP}& \multicolumn{3}{|c||}{Range in Soil}&Ratio\\
\hline
\hline
& Mass & Gamma & Used & low & high & typical & Soil \\
& Spec.& Spec. & Value &  &  &  & vs. \\
Element & $\left[\frac{\mu g}{g}\right]$
        & $\left[\frac{\mu g}{g}\right]$
        & $\left[\frac{\mu g}{g}\right]$
        & $\left[\frac{\mu g}{g}\right]$
        & $\left[\frac{\mu g}{g}\right]$
        & $\left[\frac{\mu g}{g}\right]$
        &WIPP\\
\hline
\hline
Uranium & 0.048 & $<$0.37 & 0.048 & 0.5 & 2.5 & 1.5 & 30 \\
\hline
Thorium & 0.08 & 0.25 & 0.25 & 1.2 & 3.7 & 2.4 & 10 \\
\hline
Potassium & 784 & 182 & 480 & 500 & 900 & 700 & 1.5 \\
\end{tabular} 
\end{center}
\end{table}
\begin{table}[h]
\caption{Number of events fulfilling the requirement of at least 2~MeV energy deposition 
per panel. 10$^7$ gammas and neutrons were started from a sphere of 2~m radius around the muon panels.
\label{nsim}}
\begin{center}
\begin{tabular}{r|l}
Energy & Events\\
\hline
0.5 MeV&  0 \\
1 MeV&  0 \\
2 MeV&  0 \\
3 MeV&  0 \\
4 MeV&  1\\
5 MeV&  306\\
6 MeV&  621\\
7 MeV&  814\\
8 MeV&  916\\
10 MeV& 947\\
\end{tabular}
\end{center}
\end{table}
\begin{table}[h]
\caption{The contributing parameter for the muon flux.} 
  {\label{factors}}
\begin{center}
\begin{tabular}{lr}
 Name of parameter & Contribution \\
\hline
Runtime of the experiment 				& 532800 seconds \\
Events in muon peak 					& 5224$\pm$72 Events \\ 
Contribution background to signal 		& -40$\pm$4 Events \\
Muon induced events below threshold 	& +62$\pm$6 Events \\
Neutrons from U and Th  				&  -54 Events \\
Neutrons from muons in rock				&   -2 Events \\
Gamma rays 								&   -32 Events \\
Detector Efficiency 					& 100 $\pm$ 1\% \\
Detector Efficiency (geometric)			& 84.25\% $\pm$ 0.4\% \\
Conversion factor to vertical flux 		& 0.65 $\pm$ 0.04 \\
\end{tabular} 
\end{center}
\end{table}
\clearpage

\end{document}